\documentclass[preprint,aps,prd,superscriptaddress,amssymb,amsmath,nofootinbib]{revtex4-1}

\usepackage[linktocpage,pagebackref=true]{hyperref}

\usepackage{amsmath,setspace,amsfonts,latexsym}
\usepackage{amssymb}
\usepackage{color}
\usepackage{epsfig}
\usepackage{graphicx}
\usepackage{slashed}
\usepackage{caption}
\usepackage{subcaption}

\definecolor{White}{rgb}{1,1,1}
\definecolor{Red}{rgb}{1,0.1,0}
\definecolor{LightYellow}{rgb}{1,1,.875}
\definecolor{SteelBlue}{rgb}{.273,.508,.703}
\definecolor{navy}{rgb}{0,0,.5}
\definecolor{LightCyan}{rgb}{.875,1,1}
\definecolor{DarkRed}{rgb}{.543,0,0}
\definecolor{HotPink}{rgb}{1,.41,.70}
\definecolor{ForestGreen}{rgb}{.13,.54,.13}
\definecolor{OliveDrab}{rgb}{.42,.55,.14}
\definecolor{MediumBlue}{rgb}{0,0,.80}
\definecolor{RoyalBlue}{rgb}{.25,.41,.88}
\definecolor{DeepSkyBlue}{rgb}{0,.746,1}
\definecolor{Brown}{rgb}{0.545,0.271,0.074}
\definecolor{Purple}{rgb}{0.637,0.285,0.641}

\let\vec\mathbf

\def\bea{\begin{eqnarray}}
\def\eea{\end{eqnarray}}
\def\bec{\begin{center}}
\def\ec{\end{center}}

\def\beq{\begin{equation}}
\def\eeq{\end{equation}}

\newcommand\lsim{\mathrel{\rlap{\lower4pt\hbox{\hskip1pt$\sim$}}
    \raise1pt\hbox{$<$}}}
\newcommand\gsim{\mathrel{\rlap{\lower4pt\hbox{\hskip1pt$\sim$}}
    \raise1pt\hbox{$>$}}}
\def\bea{\begin{eqnarray}}
\def\eea{\end{eqnarray}}
\def\ba{\begin{array}}
\def\ea{\end{array}}
\def\bc{\begin{center}}
\def\ec{\end{center}}

\begin{document}
\preprint{CTPU-PTC-19-02}
%

\title{\Large Light Bending in Models with a Generic Scalar Field}

\author{Dongjin Chway}
\email{djchway@gmail.com}
\affiliation{Center for Theoretical Physics of the Universe,  Institute for Basic Science, Daejeon 34051, South Korea}

\begin{abstract}

We study the deflection and time delay of light by the Sun in general scalar extensions of the Standard Model which may violate the equivalence principle.
Despite the presence of the interaction $\phi FF$ or $\phi F \tilde{F}$ between the scalar field and photon, we show that the bending and time delay of light are the same as in Einstein's general relativity.
The bending angle is obtained using geometrical optics and compared with the angle obtained using another method based on scattering amplitude.
It is pointed out that the method based on scattering amplitude can lead to wrong conclusions about potential energy and light polarization.
Also, we obtain a constraint on the generic scalar particle from the parametrized post-Newtonian parameter $\gamma$, noting that planet motions are affected by the scalar field as in scalar-tensor theories with some modifications to the scalar field's couplings to the Sun and planets.

\end{abstract}

\maketitle

\section{introduction} 

New light bosons are one of the most common remnants of a number of extensions of the Standard Model (SM)~\cite{Halverson:2018xge}.
Extra-dimensional theories such as string theory yield scalar moduli from the shape and size of the extra dimensions and pseudoscalar axions from the higher form gauge fields.
Also, in the low-energy effective perspective, new light bosons are favored: the QCD axion to explain the strong CP problem~\cite{Peccei:1977hh,Peccei:1977ur,Weinberg:1977ma,Wilczek:1977pj} and the scalar quintessence field to satisfy the de Sitter swampland conjecture~\cite{Agrawal:2018own,Obied:2018sgi,Garg:2018reu,Ooguri:2018wrx}.

Sourced by astrophysical objects, the light bosonic fields or clouds can form surrounding the objects.
For instance, a scalar Yukawa potential field can be found surrounding stars and planets if the scalar particle interacts with the SM particles.
Another example is rotating black holes producing bosonic clouds by superradiance~\cite{Damour:1976kh,Zouros:1979iw,Detweiler:1980uk,Herdeiro:2014goa,East:2017ovw,Herdeiro:2017phl}.
  
Such fields and clouds affect motions of nearby objects.
Yukawa potential field can provide an additional force between the Sun and the Earth changing the relationship between the Earth's orbital period and the mass of the Sun.
The gravitational lensing of light due to rotating black holes with bosonic clouds can lead to distinctive shadows~\cite{Cunha:2015yba,Vincent:2016sjq,Cunha:2016bjh}.
Light passing through an axion cloud bends in addition to the gravitational bending depending on its polarization~\cite{Mohanty:1993zj,Plascencia:2017kca}.

In scalar-tensor theories, the scalar field does not bend light in the classical limit. 
In the Jordan frame, it was shown in Ref.~\cite{Deng:2016moh} that the light deflection angle and time delay due to a massive object in scalar-tensor theories are the same as those given by Einstein's general relativity (GR).\footnote{This was shown also for Starobinsky gravity in Ref.~\cite{Accioly:1998nm}. A generalization of Starobinsky gravity is $f(R)$ gravity which is equivalent to scalar-tensor theories. For generalizations with derivatives, it was shown in Ref.~\cite{Giacchini:2018twk,Accioly:2016qeb} that the light deflection occurs in the same way as in GR.}
The light bending and time delay due to the bending are independent of properties of the scalar particle and dependent of $G M$, where $G$ is the gravitational constant and $M$ is the mass of the massive object.
On the other hand, orbits of planets around the massive object are determined by both the Yukawa potential produced by the scalar particle and $G M$.
Thus, information on the scalar field is extracted from planet motions combined with light motions eliminating ambiguity on $G M$.


In the Einstein frame, this is more readily seen because the scalar field interacts with the trace of the energy-momentum tensor of the photon which vanishes in the classical limit.
Furthermore, from the trace anomaly, we can expect that the scalar field might interact with photons whose energy is larger than the electron mass and deflect them.  
On the other hand, the trace of the energy-momentum tensor of non-relativistic particles (or microscopically, that of QCD) is non-zero and the scalar field mediates force between the massive object and its planets.

In this paper, we shall consider the most general singlet scalar extension of the SM without derivative couplings of the scalar field.
In general, the real scalar field, $\phi$, interacts with photons by dimension 5 operators $\phi F_{\mu \nu}F^{\mu \nu}$ and $\phi F_{\mu \nu} \tilde{F}^{\mu \nu}$.
However, we will show that the scalar field surrounding the Sun does not affect the light bending. 
In section~\ref{STT}, scalar-tensor theories in the Einstein frame are discussed in detail; especially, the Eddington parameter $\gamma$ is obtained in terms of the parameters in the Einstein frame. 
In section~\ref{GST}, the general scalar extension is introduced, and the Earth motion is discussed.
The deflection angle is obtained using geometric optics in section~\ref{GO}.
The method of computing the deflection angle from scattering amplitude is described and applied to the $\phi F_{\mu\nu} F^{\mu\nu}$ interaction in section~\ref{SAscalar} and to the $\phi F_{\mu\nu} \tilde{F}^{\mu\nu}$ interaction in section~\ref{SApseudo}.
We show that the bending via amplitude can lead to wrong conclusions about potential energy and polarization of refracted light.
Section~\ref{conclusion} is the conclusion.

\section{Scalar-Tensor Theories}
\label{STT}

In the Einstein frame, the Lagrangian density of scalar-tensor theories with a real scalar field $\phi$ can be described as 
\begin{align}
\label{lagst}
\mathcal{L} = \mathcal{L}_{\rm SM} + \frac{1}{2} \partial_\mu \phi \partial^\mu \phi -\frac{1}{2} m_\phi^2 \phi^2 - d_T \kappa \phi T_\mu^\mu
\end{align}
with traditional Einstein gravity.
Here, the inverse of the Planck mass is defined as $\kappa = (\sqrt{2} M_{\rm Pl})^{-1}= \sqrt{ 4 \pi G}$.
The trace of the energy-momentum tensor can be written as, microscopically,
\begin{align}
T_\mu^\mu = \frac{\beta_3}{2 g_3} G^A_{\mu\nu} G^{A\mu\nu}+ \frac{\beta_e}{2 e_3} F_{\mu\nu} F^{\mu\nu} +\sum_f (1 + \gamma_{m_f} ) m_f \bar{\psi}_f \psi_f
,
\end{align}
where $\beta_{3,e}$ are the beta functions of the strong and electromagnetic couplings and $\gamma_{m_f}$ is the mass anomalous dimension of a fermion $f$.
The $\beta_e$ and thus $\phi F_{\mu\nu}F^{\mu\nu}$ term vanish below the electron mass scale.
Macroscopically, a perfect fluid has
\begin{align}
T^{\mu\nu} = (\rho + p ) u^\mu u^\nu - p g^{\mu\nu},
\end{align}
where $\rho$ is the energy density, $p$ is the pressure of the fluid, $u$ is the 4-velocity of the fluid, and $g^{\mu\nu} = {\rm diag}(+1,-1,-1,-1)$ is the Minkowski metric.
For the Sun at rest with the mass, $M_{\rm S}$, the trace is 
\begin{align}
T^\mu_\mu = M_{\rm S} \delta^{(3)}(\vec{r}),
\end{align}
in the point-like approximation.
The Lagrangian \eqref{lagst} yields the equation of motion for the scalar and its solution as
\begin{align}
\label{scalareom}
\left(\partial^2 + m_\phi^2 \right) \phi + d_T \kappa M_{\rm S} \delta^{(3)}(\vec{r})  = 0,  \quad
\phi(r)  = - d_T \kappa \frac{M_{\rm S}}{4 \pi r} e^{-m_\phi r}.
\end{align}
The potential energy of the Sun and the Earth with the mass $M_{\rm E}$ is
\begin{align}
V(r_{\rm SE}) = d_T \kappa M_{\rm E} \phi(r_{\rm SE}) = - \left( d_T \kappa \right)^2 \frac{M_{\rm S}M_{\rm E}}{4 \pi r_{\rm SE}} e^{-m_\phi r_{\rm SE}} = - d_T^2  \frac{GM_{\rm S}M_{\rm E}}{r_{\rm SE}} e^{-m_\phi r_{\rm SE}} ,
\end{align}
where $r_{\rm SE}$ is the distance between the Sun and the Earth.
Therefore, the acceleration of the Earth is
\begin{align}
\label{einacc}
\frac{GM_{\rm S}}{r_{\rm SE}^2} \left( 1+ d_T^2 (1+m_\phi r_{\rm SE}) e^{-m_\phi r_{\rm SE}} \right) = \frac{ 4 \pi^2 r_{\rm SE}}{ ( 1 {\rm yr})^2}.
\end{align}
On the other hand, since the gravity in the Einstein frame is the traditional one, the light bending angle is
\begin{align}
\frac{4GM_{\rm S}}{b},
\end{align}
and the one-way time delay due to the deflected path is $2 G M \ln\left( 4 r_{\rm SE} r_{\rm Ss} / b^2 \right)$ when the source, the Sun, and the Earth are almost in a straight line with the impact parameter, $b$, and the distance between the source and the Sun, $r_{\rm Ss}$.
The time delay results in the Doppler shift to a round trip signal (Earth-spacecraft-Earth) yielding the fractional frequency shift~\cite{Bertotti1992,Iess1999}
\begin{align}
\label{einshif}
 8\frac{GM_{\rm S}}{b}\frac{\mathrm{d}b}{\mathrm{d}t}
 \end{align}
which provides the scalar-field-independent measurement for $G M_{\rm S}$ using the Cassini spacecraft~\cite{Bertotti:2003rm}.

The parametrized post-Newtonian (PPN) parameters are defined in the Jordan frame with the metric
\begin{align}
ds_{\rm Jordan}^2 = -\left(1-\frac{2 (G M_{\rm S})_{\rm J}}{\chi}\right) dt_{\rm J}^2 +  \left( 1 + \gamma \frac{2 (G M_{\rm S})_{\rm J}}{\chi} \right) \left( d\chi^2+\chi^2 d\Omega^2 \right),
\end{align}
where we keep only one relevant PPN parameter $\gamma$ which is also known as the Eddington parameter.
The Earth acceleration in this frame is
\begin{align}
\label{joracc}
\frac{(GM_{\rm S})_{\rm J}}{r_{\rm SE}^2} = \frac{ 4 \pi^2 r_{\rm SE}}{ ( 1 {\rm yr})^2}.
\end{align}
The light bending and the fractional frequency shift are, respectively,
\begin{align}
\label{jorben}
(1+\gamma)\frac{2(GM_{\rm S})_{\rm J}}{b}  ,\quad {\rm and} \quad 4(1+\gamma)\frac{(GM_{\rm S})_{\rm J}}{b}\frac{\mathrm{d}b}{\mathrm{d}t}.
\end{align}
Comparing the formulae (\ref{einacc}--\ref{einshif}, \ref{joracc}, \ref{jorben}), we obtain the Eddington parameter in terms of the parameters in the Einstein frame as
\begin{align}
\gamma = \frac{1-d_T^2 (1+m_\phi r_{\rm SE}) e^{-m_\phi r_{\rm SE}}}{1-d_T^2 (1+m_\phi r_{\rm SE}) e^{-m_\phi r_{\rm SE}}}
\end{align}
which is constrained by the Cassini tracking~\cite{Bertotti:2003rm} as $\left| \gamma - 1 \right| < 4.6 \times 10^{-3}$ at $2\sigma$.
In comparison with Ref.~\cite{Scharer:2014kya} in which the PPN parameters were computed in the Einstein frame, there are two differences; they promoted the PPN parameter $\gamma$ from a constant to a function of position $\chi$ while we keep it as a constant, and they used the closest distance (1.6 solar radii) between the signal and the Sun as the interaction distance following Ref.~\cite{Hohmann:2013rba} while we use the distance between the Sun and the Earth as the interaction distance.
The interaction distance should be $r_{\rm SE}$ because the scalar field affects planet motions, not light bending.

\section{Generic scalar theory}
\label{GST}

In the Einstein frame, after some field redefinitions, any extension of the SM with a real scalar without derivative couplings of the scalar field can be represented by the Lagrangian density~\cite{Choi:2018rze}
\begin{align}
\label{genlag}
\mathcal{L} = \mathcal{L}_{\rm SM} + \frac{1}{2} \partial_\mu \phi \partial^\mu \phi -\frac{1}{2} m_\phi^2 \phi^2 - \kappa \phi \left[ \frac{d_g \beta_3}{2 g_3} G^A_{\mu\nu} G^{A \mu\nu} + \sum_i \left( d_{m_i} + \gamma_{m_i} d_g \right) m_i \bar{\psi}_i \psi_i \right] + \mathcal{L}_{\phi \gamma \gamma},
\end{align}
where we ignore terms of $\mathcal{O}\left(\phi^3\right)$ in the scalar potential and $\mathcal{O}\left(\phi^2\right)$ in the scalar interaction with the SM.
The interaction of the scalar field with photons is provided by
\begin{align}
\label{lag}
 \mathcal{L}_{\phi \gamma\gamma} = \frac{\lambda_s}{4} \kappa \phi F_{\mu \nu} F^{\mu \nu} , \quad {\rm or} \quad  \frac{\lambda_a}{4} \kappa \phi F_{\mu \nu} \tilde{F}^{\mu \nu},
 \end{align}
 where $\tilde{F}^{\mu \nu} = (1/2) \epsilon^{\mu \nu \alpha \beta} F_{\alpha \beta}$.
 Although tests on the weak equivalence principle (WEP)~\cite{Smith:1999cr,Wagner:2012ui,Berge:2017ovy} constrain these parameters, such constraints become weaker as $d_{m_i}$ for quarks approach $d_g$, and $\lambda_s$ becomes smaller. 

Similarly to \eqref{scalareom}, the solution of the equation of motion for the scalar field is obtained as
\begin{align}
\label{phigen}
\phi(r)  = -d_{\rm S} \kappa \frac{M_{\rm S}}{4 \pi r} e^{-m_\phi r},
\end{align}
where $d_{\rm S}$ is the coupling between the scalar field and the Sun.
The coupling is determined by the constituents of the Sun as 
\begin{align}
d_{\rm S} = \sum_a d_a f_{{\rm S},a}
\end{align}
where $f_{{\rm S},a}$ is the mass fraction of an atom $a$ constituting the Sun and $d_a$ is the coupling between the scalar field and the atom $a$.
An atom $a$ of charge $Z$ and atomic number $A$ has the coupling~\cite{Damour:2010rp,Damour:2010rm}
\begin{align}
d_{a} =  d_g + \bar{\alpha}_A (d_{m_i}-d_g, \lambda_s; Z, A) ,
\end{align}
where $ \bar{\alpha}_A (d_{m_i}-d_g, \lambda_s; Z, A)$ is found in Eqs.~(17--21) of Ref.~\cite{Damour:2010rm}.
For example, the major component of the Sun is hydrogen which provides 73\% mass fraction and its coupling is
\begin{align}
d_{\rm hydrogen} = d_g + 0.048 (d_{\hat{m}}-d_g) - 0.0017 ( d_{\delta m} - d_g) + 5.5 \times 10^{-4} (d_{m_e}-d_g) + 6.7 \times 10^{-4} \lambda_s
\end{align}
with definitions $d_{\hat{m}} = \frac{ m_d d_{m_d}+m_u d_{m_u}  }{ m_d+m_u}$ and $d_{\delta m} = \frac{ m_d d_{m_d} - m_u d_{m_u} }{ m_d -m_u }$.
Similarly to \eqref{einacc}, the Earth acceleration is
\begin{align}
\label{accgen}
\frac{GM_{\rm S}}{r_{\rm SE}^2} \left( 1+ d_{\rm S} d_{\rm E} (1+m_\phi r_{\rm SE}) e^{-m_\phi r_{\rm SE}} \right) = \frac{ 4 \pi^2 r_{\rm SE}}{ ( 1 {\rm yr})^2},
\end{align}
where $d_{\rm E}$ is the coupling between the scalar field and the Earth.
Once $G M_{\rm S}$ is determined by the light bending, this can constrain the Lagrangian parameters for $m_\phi \lesssim r_{\rm SE}^{-1}$.\footnote{The Lagrangian parameters are also constrained by measuring deviation from the $1/r^2$ force~\cite{Fischbach:1996eq}. The information on the light bending is not needed, but the $1/r^2$ tests do not cover $m_\phi \ll r_{\rm SE}^{-1}$. }

\section{Geometric Optics}
\label{GO}

In this section, we apply geometric optics to obtain the light deflection angle due to the scalar field.
The equation of motion for photon from \eqref{genlag} and \eqref{lag} is
\begin{align}
\label{eom0}
\partial_\mu F^{\mu \nu} = \frac{\lambda_s \kappa}{1 - \lambda_s \kappa\phi} (\partial_\mu \phi) F^{\mu \nu} , \quad {\rm or} \quad \partial_\mu F^{\mu \nu} = \lambda_a \kappa (\partial_\mu \phi ) \tilde{F}^{\mu \nu} 
\end{align}
choosing the Lorenz gauge $\partial_\mu A^\mu = 0$.
In terms of the electric and magnetic fields, this is
\begin{align}
\label{eom1}
\vec{\nabla} \cdot \vec{E} &= \frac{\lambda_s \kappa }{1 - \lambda_s  \kappa \phi} \vec{\nabla} \phi \cdot \vec{E} 
\quad\qquad \qquad, {\rm or} & \qquad \vec{\nabla} \cdot \vec{E} &=\lambda_a \kappa \vec{\nabla}\phi \cdot \vec{B} \nonumber \\
\vec{\nabla} \times \vec{B} -\frac{\partial \vec{E}}{\partial t}&= \frac{\lambda_s \kappa }{1 - \lambda_s \kappa \phi}  \left( \vec{\nabla}\phi \times \vec{B} -\frac{\partial \phi }{\partial t}  \vec{E}\right)
& \vec{\nabla} \times \vec{B} -\frac{\partial \vec{E}}{\partial t}&= \lambda_a \kappa \left( -\vec{\nabla}\phi \times \vec{E} - \frac{\partial \phi }{\partial t}\vec{B} \right)
. 
\end{align}
The Bianchi identity
\begin{align}
\partial_\mu \tilde{F}^{\mu\nu} = 0
\end{align}
provides 
\begin{align}
\label{eom2}
\vec{\nabla} \times \vec{E} + \frac{ \partial \vec{B}}{\partial t} = 0,
\\
\vec{\nabla} \cdot \vec{B} = 0.
\nonumber
\end{align}
From \eqref{eom1} and \eqref{eom2}, we obtain wave equations
\begin{align}
\label{wave}
(\partial_t^2 - \vec{\nabla}^2) \vec{E} &= \frac{\lambda_s \kappa }{1 - \lambda_s \kappa \phi} \partial_\mu \phi \partial^\mu \vec{E} 
\quad \qquad, {\rm or} & (\partial_t^2 - \vec{\nabla}^2) \vec{E} &= \lambda_a \kappa \partial_\mu \phi \partial^\mu \vec{B}  \\
(\partial_t^2 - \vec{\nabla}^2) \vec{B} &= \frac{\lambda_s \kappa }{1 - \lambda_s \kappa \phi} \partial_\mu \phi \partial^\mu \vec{B} 
&(\partial_t^2 - \vec{\nabla}^2) \vec{B} &= - \lambda_a \kappa \partial_\mu \phi \partial^\mu \vec{E}  
. \nonumber
\end{align}
in the small wavelength limit where the wavelength of the photon is much smaller than the characteristic length of the scalar field $\phi$; in other words, we neglect terms with $\partial^2 \phi$ or $\partial^\mu \phi \partial^\nu \phi$.
Therefore, the dispersion relation is 
\begin{align}
k^2-\frac{\lambda_s \kappa }{1- \lambda_s \kappa \phi} i k^\mu \partial_\mu \phi =0, \quad {\rm or} \quad k^2 \pm \lambda_a \kappa  k^\mu \partial_\mu \phi  = 0,
\end{align}
where the sign $\pm$ corresponds to the right- and left-circular polarization; we make an ansatz $\vec{E} ( t,\vec{x}) \propto (\hat{\epsilon}_1 \pm i \hat{\epsilon}_2) e^{i k \cdot x}$ and $\vec{B} ( t,\vec{x}) \propto (\hat{\epsilon}_2 \mp i \hat{\epsilon}_1) e^{i k \cdot x}$ with $\hat{\epsilon}_{1} \perp \hat{\epsilon}_{2}$ and $\hat{\epsilon}_1 \times \hat{\epsilon}_2 \parallel \vec{k}$.

The index of refraction is obtained from the dispersion relation as
\begin{align}
\label{indexofref}
n=\frac{ |\vec{k}|}{k_0} \simeq 1 -  i \frac{ \lambda_s \kappa}{2(1-\lambda_s \kappa \phi)} \left( \frac{1}{k_0} \frac{ \partial \phi}{ \partial t} + \frac{\vec{k} \cdot \vec{\nabla} \phi}{ k_0^2} \right)
,\,\, {\rm or} \quad
1 \pm \frac{\lambda_a \kappa}{2}  \left( \frac{1}{k_0} \frac{ \partial \phi}{ \partial t} + \frac{\vec{k} \cdot \vec{\nabla} \phi}{ k_0^2} \right).
\end{align}
It is noteworthy that the real part of the index of refraction for the $\phi F_{\mu \nu} F^{\mu \nu}$ interaction is $1$ in the $\lambda_s^1$ order. Thus, the scalar field does not change the direction of the photon propagation. This can be more promptly seen by observing, in \eqref{eom1}, that the electric permeability is $\epsilon = 1 -\lambda_s \kappa \phi$, the magnetic permeability is $\mu=1/(1-\lambda_s \kappa \phi)$, and therefore, the refraction index is $n=\sqrt{ \epsilon \mu} = 1$.

If the spacetime is curved, it provides an additional contribution to the index of refraction.
For a spherically symmetric metric
\begin{align}
ds^2 = A(r) dt^2 -B(r) dr^2 - r^2 d \Omega^2,
\end{align}
the additional index of refraction due to the gravity,
\begin{align}
\Delta n_{\rm gravity} = \sqrt{\frac{B(r)}{A(r)} }-1,
\end{align}
should be added to \eqref{indexofref}. 
The Schwarzschild metric gives $ \Delta n_{\rm gravity} \simeq 2 G M /r$.

Given the index of refraction, the trajectory of light, $\vec{r}(s)$, is obtained by
\begin{align}
\frac{d}{ds} n \frac{d \vec{r}}{ds} = \vec{\nabla} n,
\end{align}
where $ds = dt/n$ is the path length.
As a zeroth order solution, the trajectory is
\begin{align}
\vec{r}_0(t) = b \hat{x} + t \hat{y}, \qquad -\infty < t < \infty,
\end{align}
where $b$ is the impact parameter.
The deflection angle, $\theta$, is
\begin{align}
\label{thetadef}
\theta \simeq \sin \theta = - \hat{x} \cdot \int_{-\infty}^{\infty} \frac{d^2\vec{r}}{dt^2} dt \simeq  - \int_{-\infty}^{\infty}  \hat{x} \cdot \vec{\nabla} n (\vec{r}_0)  \,\, dt .
\end{align}
For example, the Schwarzschild metric yields
\begin{align}
\theta_{\rm gravity} = \int_{-\infty}^\infty \frac{2 G M b}{(b^2 + t^2)^{3/2}} dt = \frac{4 G M}{b}.
\end{align}
The deflection angle due to the scalar field is $\theta_{\rm scalar}=0$ for the $\phi F_{\mu\nu} F^{\mu\nu}$ interaction and
\begin{align}
\label{pseutheta}
\theta_{\rm scalar} = \mp \frac{\lambda_a \kappa}{2 k_0 } \int_{-\infty}^{\infty} \left. \frac{\partial }{\partial x}  \left( \frac{ \partial \phi}{ \partial t} + \frac{\partial \phi}{\partial y} \right) \right|_{\vec{r}_0(t)} dt
\end{align}
for the $\phi F_{\mu\nu} \tilde{F}^{\mu \nu}$ interaction.
This can be further simplified to $\theta_{\rm scalar} = 0$ if the scalar field satisfies $\frac{\partial \phi}{\partial t}=0$ and $\frac{\partial \phi}{\partial y} \left(\vec{r}_0(t)\right)=-\frac{\partial \phi}{\partial y} \left(\vec{r}_0(-t)\right)$; for example, $\phi = \phi(r)$ does not bend light.
Thus, the scalar field \eqref{phigen} surrounding the Sun does not bend light at this order.

The light bending is purely gravitational.\footnote{Even if the scalar field depends on time and bends light, its contribution is suppressed by $e^{-m_\phi l_\phi}/(k_0 l_\phi) $ where $l_\phi$ is the characteristic length scale of the scalar field. In \eqref{eddingtonp}, $\lambda_{s,a} e^{-m_\phi l_\phi}/(k_0 l_\phi) $ would have been added to $d_{\rm E}e^{-m_\phi r_{\rm SE}}$.} 
Combined with the Earth acceleration \eqref{accgen}, the Eddington parameter in general scalar extensions of the SM is
\begin{align}
\label{eddingtonp}
\gamma = \frac{1-d_{\rm S} d_{\rm E} (1+m_\phi r_{\rm SE}) e^{-m_\phi r_{\rm SE}}}{1-d_{\rm S} d_{\rm E} (1+m_\phi r_{\rm SE}) e^{-m_\phi r_{\rm SE}}}.
\end{align}
The measurements on the WEP~\cite{Smith:1999cr,Wagner:2012ui,Berge:2017ovy} provide stronger constraints on the scalar couplings as~\cite{Choi:2018rze}
\begin{align}
d_g \lesssim 3 \times 10^{-6}, \quad d_{m_q} \lesssim 10^{-5}, \quad \lambda_s \lesssim 2 \times 10^{-4} \quad {\rm for} \,\, m_\phi \lesssim 10^{-13}{\rm eV}
\end{align}
when there is no special cancellation among the couplings.
However, when $d_g - d_{m_q}$ and $\lambda_s$ are tiny, the WEP tests do not limit $d_g$ and $d_{m_q}$ while the constraint on the PPN parameter $\gamma$~\cite{Bertotti:2003rm} constrains them; for example, $d_g \lesssim 3 \times 10^{-3}$ when $d_g = d_{m_q}$ and $m_\phi < r_{\rm SE}^{-1}$.

\section{Scattering Amplitude by $\phi FF$}
\label{SAscalar}

Given the non-zero coupling between the scalar field and photon, the amplitude of photon scattering does not vanish in general, which might sound inconsistent with the non-bending of light obtained in the previous section. 
In this section, we use a semiclassical approach of obtaining light bending from the scattering amplitude to understand the non-bending for the $\phi F_{\mu\nu} F^{\mu\nu}$ interaction, emphasizing an ambiguity in the process.
The bending via amplitudes has been used to compute the leading quantum correction to the gravitational bending of light~\cite{Bjerrum-Bohr:2014zsa,Donoghue:2015xla,Bjerrum-Bohr:2016hpa}.

Classical potential energy is related with the scattering amplitude $\mathcal{M}$ by the Fourier transformation
\begin{align}
V(\vec{r}) = \frac{1}{4 M k_0} \int \mathcal{M}(\vec{q}) e^{i \vec{q} \cdot \vec{r}} \frac{ d^3 q}{(2 \pi)^3},
\end{align}
in the small momentum transfer limit.
For example, the potential energy between the Sun and the Earth is
\begin{align}
V_{\rm SE}(r) &= \frac{1}{4 M_{\rm S} M_{\rm E}}\int  i (- 2 i d_{\rm S} \kappa M_{\rm S}^2) \frac{i }{(k-k')^2 - m_\phi^2} (-2 i d_{\rm E} \kappa M_{\rm E}^2) e^{i \vec{q} \cdot \vec{r}} \frac{ d^3 q}{(2 \pi)^3} \\
&\simeq - d_{\rm S} d_{\rm E}  \frac{G M_{\rm S} M_{\rm E}}{r} e^{- m_\phi r}
,
\end{align}
where $k$ ($k'$) is the incoming (outgoing) momentum of the scattered object which is the Earth in this case, and $(k'-k)^\mu = (0, \vec{q})$. 
For the light scattering by $\phi F_{\mu\nu}F^{\mu\nu}$, the scattering amplitude is
\begin{align}
\mathcal{M}_{\rm scalar} = i (- 2 i d_{\rm S} \kappa M_{\rm S}^2) i \frac{(k \cdot k') (\epsilon \cdot \epsilon'^* )- (k \cdot \epsilon'^* )(k' \cdot \epsilon ) }{(k-k')^2 - m_\phi^2} ( i \lambda_s \kappa)
.
\end{align}
In order to analyze the polarization dependence, we set the photon momenta and photon polarizations as
\begin{align}
k^\mu &= k_0 ( 1, 0, 1, 0) , \quad& k'^\mu &= k_0 (1, - \sin \Theta, \cos \Theta, 0) \\
\epsilon_z^\mu &= ( 0, 0, 0, 1) , \quad& \epsilon_z'^\mu &= ( 0, 0, 0, 1)\\
\epsilon_x^\mu &= ( 0, 1, 0, 0) , \quad& \epsilon_x'^\mu &= ( 0, \cos \Theta, \sin \Theta, 0)\\
\epsilon_\pm^\mu &= \frac{1}{\sqrt{2}} ( 0, \pm i, 0, 1) , \quad& \epsilon_\pm'^\mu &= \frac{1}{\sqrt{2}} ( 0, \pm i \cos \Theta, \pm i \sin \Theta, 1).
\end{align}
Here, the angular parameter $\Theta$ of the phase space should not be confused with the final bending angle $\theta$ in \eqref{thetadef}.
The bending angle $\theta$ is
\begin{align}
\theta \simeq \sin \theta = - \hat{x} \cdot \int_{-\infty}^{\infty} \frac{d^2\vec{r}}{dt^2} dt \simeq  \frac{1}{k_0} \int_{-\infty}^{\infty}  \hat{x} \cdot \vec{\nabla} V (\vec{r}_0)  \,\, dt .
\end{align}
We obtain
\begin{align}
 (k \cdot k') (\epsilon \cdot \epsilon'^* )- (k \cdot \epsilon'^* )(k' \cdot \epsilon ) = 
   \begin{cases}
    -k_0^2 (1 - \cos \Theta) &\epsilon_z \rightarrow \epsilon'_z, {\rm or} \, \epsilon_\pm \rightarrow \epsilon'_\mp \\
    k_0^2 (1 - \cos \Theta) &\epsilon_x \rightarrow \epsilon'_x\\
   0 &\epsilon_x \rightarrow \epsilon'_z, \epsilon_z \rightarrow \epsilon'_x, {\rm or} \,   \epsilon_\pm \rightarrow \epsilon'_\pm
   \end{cases}
   \end{align}
   and the potential energy is 
   \begin{align}
V(\vec{r})
&= \frac{1}{2k_0} \int  (-  d_{\rm S} \kappa M_{\rm S})  \frac{\mp k_0^2 (1-\cos\Theta) }{-\vec{q}^2 - m_\phi^2} ( \lambda_s \kappa)  e^{i \vec{q} \cdot \vec{r}} \frac{d^3 q}{(2\pi)^3} .
\end{align}
It is important to note that $\Theta$ as a function of $\vec{q}$ has an ambiguity as
\begin{align}
k \cdot (k' - k) = - \vec{k} \cdot \vec{q} = k_0^2 ( 1 - \cos \Theta ) , \quad {\rm and} \quad
(k'-k)^2 = - \vec{q}^2 = -2 k_0^2 ( 1 - \cos \Theta).
\end{align}
Choosing the first expression, the potential energy is
\begin{align}
V(\vec{r}) &=    \frac{ \lambda_s  d_{\rm S} \kappa^2 M_{\rm S}}{2k_0} \int \frac{\pm \vec{k} \cdot \vec{q}  }{\vec{q}^2 + m_\phi^2}  e^{i \vec{q} \cdot \vec{r}} \frac{d^3 q}{(2\pi)^3} \\
&=  \mp i \frac{ \lambda_s  d_{\rm S} \kappa^2 M_{\rm S} }{2 k_0} \vec{k} \cdot \vec{\nabla} \frac{1}{4\pi r} e^{-m_\phi r}
\end{align}
which is purely imaginary and irrelevant to bending.
Contrarily, the second expression provides
\begin{align}
V(\vec{r}) &=    \frac{ \lambda_s  d_{\rm S} \kappa^2 M_{\rm S} }{2 k_0} \int \frac{\mp  \vec{q}^2/2  }{\vec{q}^2 + m_\phi^2}  e^{i \vec{q} \cdot \vec{r}} \frac{d^3 q}{(2\pi)^3} \\
&=  \mp \frac{ \lambda_s  d_{\rm S} \kappa^2 M_{\rm S} }{2k_0} \left( \frac{\delta^{(3)}(\vec{r})}{2} - \frac{m_\phi^2}{2} \frac{1}{4\pi r} e^{-m_\phi r} \right)
\end{align}
whose second term yields non-zero bending.
Unlike in the case of Eq.~(7.5) in Ref.~\cite{Donoghue:2015xla}, the first choice $\cos \Theta = 1+ \vec{k} \cdot \vec{q} / k_0^2$ should be used here.


\section{Scattering Amplitude by $\phi F\tilde{F}$}
\label{SApseudo}
In this section, we consider the $\phi F_{\mu \nu} \tilde{F}^{\mu\nu}$ interaction which provides the scattering amplitude
\begin{align}
\mathcal{M}_{\rm scalar} = i (-2 i d_{\rm S} \kappa M_{\rm S}^2) i \frac{ \epsilon^{\mu\nu\rho\sigma}k_\mu \epsilon_\nu k'_\rho \epsilon'^*_\sigma }{(k-k')^2 - m_\phi^2} (  i \lambda_a \kappa),
\end{align}
with
\begin{align}
 \epsilon^{\mu\nu\rho\sigma}k_\mu \epsilon_\nu k'_\rho \epsilon'^*_\sigma= 
   \begin{cases}
    i k_0^2 (1 - \cos \Theta) & \epsilon_+ \rightarrow \epsilon'_- \\
    - i k_0^2 (1 - \cos \Theta) & \epsilon_- \rightarrow \epsilon'_+ \\
    k_0^2 (1 - \cos \Theta) &\epsilon_x \rightarrow \epsilon'_z , {\rm or} \,  \epsilon_z \rightarrow \epsilon'_x \\
   0 & \epsilon_\pm \rightarrow \epsilon'_\pm, \epsilon_x \rightarrow \epsilon'_x, {\rm or} \,  \epsilon_z \rightarrow \epsilon'_z
   \end{cases} .
   \end{align}
Choosing, again, $\cos \Theta = 1+ \vec{k} \cdot \vec{q} / k_0^2$ instead of $\cos \Theta = 1- \vec{q}^2 / (2k_0^2)$, the potential energy for flipping circularly polarized photon $\epsilon_\pm \rightarrow \epsilon'_\mp$ is obtained as
 \begin{align}
V(\vec{r}) 
&= \frac{1}{2k_0} \int  (-  d_{\rm S} \kappa M_{\rm S})  \frac{\pm i k_0^2 (1-\cos\Theta) }{-\vec{q}^2 - m_\phi^2} ( \lambda_a \kappa)  e^{i \vec{q} \cdot \vec{r}} \frac{d^3 q}{(2\pi)^3}  \\
&=    \frac{ \lambda_a  d_{\rm S} \kappa^2 M_{\rm S}}{2k_0} \int   \frac{\mp i \vec{k} \cdot \vec{q}  }{\vec{q}^2 + m_\phi^2}  e^{i \vec{q} \cdot \vec{r}} \frac{d^3 q}{(2\pi)^3} \\
&=  \mp  \frac{ \lambda_a  d_{\rm S} \kappa^2 M_{\rm S}}{2k_0} \vec{k} \cdot \vec{\nabla} \frac{1}{4\pi r} e^{-m_\phi r}.
\end{align}
Therefore, the deflection angle is
\begin{align}
\theta \simeq    \frac{1}{k_0} \int_{-\infty}^{\infty}  \hat{x} \cdot \vec{\nabla} V (\vec{r}_0)  \,\, dt = \mp \frac{\lambda_a \kappa}{2k_0} \int_{-\infty}^{\infty} \left. \frac{\partial}{\partial x} \frac{\partial}{\partial y} \phi(r)  \right|_{\vec{r}_0(t)}dt,
\end{align}
where $\phi(r)$ is given in \eqref{phigen}. This is the same as the bending angle \eqref{pseutheta} obtained using geometrical optics.
However, this angle is obtained from the helicity-flip amplitude while \eqref{pseutheta} is for helicity non-flip. 
The potential energies for the linearly polarized photons are even more unclear; they are either purely imaginary or zero.

The equation of motion \eqref{eom0} for the linearly polarized photon is
\begin{align}
\begin{bmatrix}
-\partial^2      &   2 \lambda_a (i k_0) (\partial_0 \phi + \partial_2 \phi)             \\
- 2 \lambda_a (i k_0) (\partial_0 \phi + \partial_2 \phi)      &   -\partial^2               \\
            \end{bmatrix}
\begin{bmatrix}
A_1     \\
A_3     \\
            \end{bmatrix}
   =
0,
\end{align}
where we choose gauge $A_0=0$ and $k \cdot A =0$ (for $\vec{k} \parallel \hat{y}$, $A_2=0$).
The photon propagation is obtained by diagonalizing the matrix and solving differential equations.
The helicity flip occurs if the diagonalization is spacetime dependent.
However, the above matrix is diagonalized by
\begin{align}
\begin{bmatrix}
A_+     \\
A_-    \\
            \end{bmatrix}
            = \frac{1}{\sqrt{2}}
            \begin{bmatrix}
i     &  1            \\
-i    &   1         \\
            \end{bmatrix}
\begin{bmatrix}
A_1     \\
A_3    \\
            \end{bmatrix}
    \end{align}
as
\begin{align}
\begin{bmatrix}
-\partial^2 -2 \lambda_a ( k_0) (\partial_0 \phi + \partial_2 \phi)      &   0            \\
0     &   -\partial^2 +2 \lambda_a ( k_0) (\partial_0 \phi + \partial_2 \phi)              \\
            \end{bmatrix}
\begin{bmatrix}
A_+     \\
A_-     \\
            \end{bmatrix}
   =
0.
\end{align}
Therefore, helicity does not flip during the propagation. The method of obtaining deflection angle based on scattering amplitude should be taken with a grain of salt.




\section{\label{conclusion} conclusion}
 
In this paper we examined the light bending by the Sun in extensions of the SM with a generic scalar field that can violate the weak equivalence principle.
Using geometric optics, we showed that the scalar field surrounding the Sun does not bend light even if the Lagrangian density has interaction terms between the scalar field and photon.
The light bending by the Sun and Shapiro time delay due to the bending are purely by gravity.
Measuring those gravitational effects constrains the scalar particle's couplings to the SM, combined with the effect of the scalar particle on the force between the Sun and its planets.
In terms of the couplings and the scalar mass, we represented the PPN parameter $\gamma$ which is constrained in \cite{Bertotti:2003rm} as $\left| \gamma - 1 \right| < 4.6 \times 10^{-3}$ at $2\sigma$.

We displayed how the non-bending can be consistent with the non-zero scattering amplitude produced by the coupling between the scalar field and photon.
The potential energy of light was extracted from the amplitude, and used for computing the force on the light. 
However, we encountered two difficulties in this method.
First, interpreting the phase space parameter into the momentum transfer was not uniquely determined.
Second, the polarization-flip amplitude produced the bending formula of the polarization-non-flip propagation.

\begin{acknowledgments}
This work was supported by IBS under the project code IBS-R018-D1. I thank Chang Sub Shin for the useful discussion and for drawing my attention to the work of Xue-Mei Deng and Yi Xie. 
\end{acknowledgments}


\end{document}